\documentstyle[aps,epsf]{revtex}
\begin{document}

\title{Strong interconversion of non-polar phonons
and Josephson plasma oscillations induced by equilibrium
Josephson currents in high
$T_c$ superconductors}
\author{A. B. Kuklov, V. S. Boyko}
\address{ Department of Engineering 
Science and Physics,
The College of  Staten Island, CUNY,
     Staten Island, NY 10314}

\maketitle

\begin{abstract}
We analyze consequences of dynamical modulations
of Josephson current by non-polar lattice mode 
in the Josephson junction barrier.
In the high $T_c$
junctions, the effect of
such modulations can be anomalously strong due to the proximity
of the insulating barrier to the superconducting state.
Accordingly, the interconversion of
sound (as well as other non-polar phonons) and the Josephson plasma 
oscillations mediated
by stationary Josephson currents, which may be present in the
junction due to various reasons, becomes possible. 
We suggest that this effect can be employed
for imaging of the stationary 
Josephson currents. Estimates of the effect 
are given.
\\


\end{abstract}\vskip0.5 cm 

\section{Introduction}

The nature of the superconductivity 
in the high $T_c$ materials continues to be
a subject of hot debates. Following
the suggestion \cite{SIG}, very significant
advances have been made in revealing the
dominant d-wave symmetry of the superconducting
order parameter (OP) in these materials \cite{DWAVE}.
A direct consequence of this symmetry is 
the possibility of the
$\pi$- Josephson coupling 
\cite{BUL}.
The existence of the $\pi$-type Josephson
junction (JJ) and feasibility of its controlled
implementation in circuits have been firmly
established \cite{DWAVE,PICIRC}. The most
startling consequence of the $\pi$-type
coupling is the spontaneous Josephson current
which generates the magnetic flux
characterized by the half of the standard unit.
It should be noted that a strong interest
to the unconventional JJ made of the high-$T_c$
materials is inspired, besides purely academic   
attention, by the very recent proposal \cite{QUBIT}
of employing such JJ as a qubit for quantum
computations. 

Another intriguing consequence of the non-traditional
symmetry of the OP
is the time-reversal symmetry breaking (TRSB)
at the grain boundaries \cite{TRBS}. 
An automatic consequence
of such states are spontaneous currents
leading to fractional vortices, characterized
by the magnetic flux which is neither the half
nor the whole unit. It is worth noting
that spontaneous fractional vortices have been 
detected in recent
years \cite{FVORTEX}. 
Their exact nature, however, is still alluding
the explanation. 

Detection of single vortices is usually achieved by
means of the Scanning SQUID Microscopy
technique (see in Ref.\cite{SQUID}). It relies on
measuring the total magnetic flux threading the pick up
loop which scans the surface from the distance of few 
$\mu$m. Thus, only those spontaneous currents can be revealed
which produce magnetic field 
normal to the surface. In fact, the resolution of the method 
\cite{SQUID} is limited by the pick up loop size
(4-18$\mu$m) as well as by the distance between
the loop and the surface. 
Furthermore, the inductive
interaction between the loop and the fluxes
may distort the actual current distribution.  
In this regard we note 
that it would be desirable to have a 
technique capable of a direct high resolution
imaging of the
spatial distribution of the stationary (spontaneous) Josephson 
currents regardless of their orientation and of
the scale of their spatial variations.

In this paper, we discuss the 
effect which could  be employed for  
the imaging of the Josephson currents with high resolution.
This effect is based on the
empirical observations of a very strong sensitivity of 
values of the Josephson critical current density ~$J_c$~ 
in the 
grain boundary (GB) JJ to the 
misorientation angle between the grains
\cite{GB}. 
The nature of this effect remains very controversial.
Many explanations have been proposed.
One of them relies on the d-wave symmetry of the OP and
faceting  of the boundary, so that the macroscopic 
(average) critical current density 
is much smaller than the local 
critical current density \cite{GB1}, which may change 
sign due to the $\pi$-type coupling.
As suggested in Ref.\cite{GB2}, an extremely
strong suppression of ~$J_c$~ is a result 
of mechanical deformations concentrated at the GB.
It has been suggested that superconductivity is
suppressed locally as long as the local
mechanical strain becomes
$\geq$ 1\% \cite{GB2}. In accordance with
the phenomenological approach \cite{GB3}, 
the strain shifts the chemical potential of the JJ 
barrier, and this may strongly affect the Josephson
coupling, as long as the barrier is characterized
by the proximity to the metal-insulator transition. 
Many properties
of the GB JJ have been successfully explained 
by this model. 
On the microscopic level, in Ref.\cite{VALENCE}
it has been pointed out 
that small changes
of the local structure, existing in the 
vicinity of the GB, should
induce large variations of the valence
of the copper ions, which control the 
free charge carriers concentration. 

Recently, the anomalous proximity effect
has been induced by light
in the Y-Ba-Cu-O JJ \cite{PROX}.  This indicates that 
the JJ barrier can easily be perturbed 
(in this case, by light)
into the superconducting state. As suggested
in Ref.\cite{QCL}, this effect can naturally
be explained if the electron system
in the barrier is a sort of {\it nearly superconducting
quantum critical liquid}, so that 
any external factor can easily shift
its equilibrium. In fact, such a factor could be - light;
mechanical deformations; optical displacements
of atoms, etc. 

In this paper, we pose a general question about
mechanisms by which deformations (lattice displacements,
in general)
affect the Josephson coupling in the GB JJ made
of the high-$T_c$ materials: 

{\it Is the role of the
deformations primary or secondary?} 
\\
\noindent
If the effect of deformations
is primary with respect to the electronic
system, it can be revealed 
by means of applying reasonable deformations externally 
in the superconducting
state and observing subsequent strong changes of ~$J_c$.
If it is secondary, 
the deformations may not affect directly the electronic
system. These may predetermine the electronic properties
during the formation of the GB JJ at high
temperatures (way above the superconducting temperatures)
through the processes of, e.g., oxygen depletion controlled
by local mechanical stress and oxygen diffusion, so that
no significant
direct effect due to the external 
deformations should be expected 
to occur in the superconducting phase. 
In order to resolve this problem we note
that, if the deformations play the primary
role, a strong {\it dynamical} coupling
should exist between sound and the Josephson
phase. In contrast, if the deformations play
a secondary role, no such a significant dynamical
coupling should be observed. 

In this paper we will assume that the
deformations play the {\it primary } role
in accordance with the general concepts 
\cite{GB3,QCL}, and will
concentrate on 
the dynamical implications of the 
strong sensitivity of ~$J_c$~ to the mechanical and,
possibly, optical (non-polar) deformations of the 
crystal lattice. Specifically,
we will analyze a coupled dynamics of
the phonons and the Josephson phase.
We suggest that the coupling between
the phase and the phonons should be anomalously 
large because of the proximity of
the barrier to the superconducting state.
In this phenomenological analysis,
no quasi-particle effects are considered. 
This approximation
can be well
justified by the requirement that the phonon
frequencies under consideration
are much smaller than a typical
superconducting
gap. In this sense, we will be
employing an adiabatic approximation
in which the superconducting electronic ensemble
adjusts itself momentarily to 
Josephson phase and the phonon variable(s). 
Thus, the following consideration is limited
by the framework of the coupled dynamical Josephson
equation \cite{BARONE} and equation
of motion for 
the phonon variable.
 
It is important to emphasize, that the
first order coupling between small 
oscillations of the Josephson phase
and sound (or optical non-polar phonons)
can exist in a medium characterized 
by the inversion symmetry, 
if only some equilibrium
Josephson currents are present. 
In other words, the linear coupling between
the phase and the non-polar phonons
can exist if and only if the time reversal
symmetry is broken (either spontaneously
or externally).  
This 
contrasts with the case of the polar
phonons, which are coupled to 
the phase oscillations through the
dynamical electric fields (due to the Josephson
relation) regardless of the presence 
of the equilibrium currents. 
In this respect, we limit our consideration
by the sound and the non-polar optical phonons, 
so that 
a mere presence of the linear coupling
between the phase and the non-polar phonons
is indicative of the 
stationary Josephson currents
(we do not consider the case
when no inversion symmetry exists).
We believe that
predictions following from our
analysis provide an opportunity for
experimental testing of
the dynamical properties of the JJ barrier
in the high-$T_c$ materials, and for 
answering
fundamental questions about the
nature of the Josephson effect in the
high-$T_c$ materials. 
Furthermore,
the discussed effect
could be 
employed for the imaging of the stationary
Josephson currents. This, in its turn, may
shed light on the nature of the spontaneous
Josephson currents and the TRSB in these
materials. 

In Sec.II, we show how the linear coupling
between the Josephson phase and the non-polar
phonons is induced by the stationary Josephson
currents. Then, the implications of this coupling
are considered. Specifically,
in Sec.III, the effect of the conversion
of the non-polar phonons into the phase oscillations
are discussed. 
In Sec. IV, we consider the induced infra-red (IR) 
activity of the non-polar optical phonons
in the JJ. Then, in Sec. V, the effect of the coherent
generation of sound in the JJ by the incoming
microwave radiation is analyzed, and the 
possibility of employing this effect for 
the imaging of the currents is addressed. 
The role of external magnetic field in enhancing
the interconversion is discussed in Sec. VI.
Finally, the discussion and conclusion are given. 

\section{ Coupling between the Josephson phase
and non-polar phonons}

The Josephson current density ~$J(\varphi)$~
as a function
of the phase difference ~$\varphi$~ can be
sensitive to atomic displacements in
the barrier.
Let us assume that
~$\xi({\bf x}, t)$~ is some phonon variable. It
can be the strain tensor ~$u_{ij}$,
or some other variable describing optical
excitations. 
If this 
variable does not correspond to a polar (IR-active )
excitation, a direct coupling between 
~$\xi$~ and the phase oscillations is forbidden.
In what follows, we will show how
this non-polar phonon can become IR-active in
the presence of the stationary
Josephson currents, so 
that
monitoring the IR absorption at
the frequency of the crystal
mode, which is known to be not IR-active 
in the absence of the static currents,  
would provide some information about these currents.

In our approach, the Josephson equation
for the phase ~$\varphi ({\bf x}, t)$~ 
contains terms 
~$\sim \xi({\bf x}, t)$. 
If ~$\xi$, which modifies somehow the 
tunneling matrix element between the sides
of the JJ, is small, the Josephson coupling energy
density (per unit area of the junction)
 ~$E_J=E_J(\varphi  , \xi)$~
can be expanded in ~$\xi$.
The symmetry consideration may require
that no linear term ~$\sim \xi$~ is
present in the ideal lattice. This is exactly the
case for, e.g., the non-polar vector and quadru-polar 
modes. If, however, ~$\xi$~ is the strain, the
scalar combination ~$ \xi=u_{jj}$~ (we employ the
convention of the summation over the repeated
indices) is allowed
in the expansion. It should also be taken into
account that the crystal structure close to the GB
and at the GB can be highly distorted.
For example, some stationary strain
~$u^{(0)}_{ij}$~ is inevitably present in any
GB junction just to ensure matching
of the sides. This, relaxes the symmetry requirement,
so that the linear term becomes allowed. 
Thus, the expansion
has a form 

\begin{eqnarray}
\displaystyle
E_J=E_J^{(0)}(\varphi)[1 + b\xi({\bf x}, t)
+o(\xi^2)],
\label{1}
\end{eqnarray}
\noindent
where ~$ E_J^{(0)} (\varphi)$~ denotes 
the Josephson energy density as a function
of the phase in the absence of ~$\xi$;
the coefficient ~$b$~ is taken
as some constant. In fact, the observations
\cite{GB,GB2}
of the very strong sensitivity of the
critical current to the misorientation angle,
which could be due to the local
deformations arising in order to insure
matching of the sides,
justify the
values as large as ~$|b|\approx 10^2$. Indeed,
it has been suggested that the deformations 
~$u_0$~ as small
as 1\% may cause the 
reduction of the current
by almost an order of magnitude \cite{GB2}. 
Thus, we take
~$|b|\approx 1/u_0 \approx 10^2$. 
As we mentioned above, the optical 
deformations of the lattice may produce 
a similar effect. In this case,
 the estimate 
for ~$b$~ can be obtained from the following
considerations: optical deformation ~$\xi$~ of 
~$1\%$~ of the lattice cell (of the size 
~$a \approx 0.4$nm)
may produce
the same effect as the strain, thus ~$|b|\approx 10^2/a$.
It should also be noted that the form
(\ref{1}) implies that the phonons do not
modify the phase dependence of the JJ energy.

The Josephson current, then, follows from
eq.(\ref{1}) as \cite{BARONE}

\begin{eqnarray}
\displaystyle
J(\varphi ,\xi)= { 2e \over \hbar }
{\partial E_J\over \partial \varphi}
=J_0(\varphi)[1 + b\xi({\bf x}, t) + o(\xi^2)],
\label{1_1}
\end{eqnarray}
\noindent
where ~$J_0(\varphi) =(2e/\hbar)\partial E^{(0)}_J/
 \partial \varphi $~ is the Josephson current
in the absence of ~$\xi$, with ~$e >0$~ being
the unit charge.

The total Josephson energy of the system is
 obtained as the integral 
~$H_J=\int d^2x E_J$~
over the area of the JJ. The 
energy ~$H_{\xi}$~ of the field ~$\xi$~
can be represented in the spirit 
of a general Landau expansion with respect
to the smallness of ~$\xi$~ as

\begin{eqnarray}
\displaystyle
H_{\xi}=\int d^3x[{\rho \dot{\xi}^2\over 2} 
+ {\rho \omega_0^2\over 2} \xi^2 +
{\rho v_g^2\over 2} (\nabla \xi)^2 + o(\xi^4)] ,
\label{101}  
\end{eqnarray}
\noindent
 where ~$\rho, \,\,\, \omega_0,\,\, v_g$~
are phenomenological constants, so that
~$\rho$~ carries the meaning of some effective
mass density. Accordingly, ~$\omega_0$~ 
and ~$v_g$~ determine the 
dispersion law of small oscillations
of ~$\xi$. 

The total energy ~$H$~ is the sum of ~$H_{\xi}+H_J$~
and of the electro-magnetic energy ~$H_{EM}=
\int d^2x[CV^2/2 + \kappa ({\bf \nabla}\varphi)^2/2]$~
of the JJ \cite{BARONE},
where ~$C$~ stands for the electric capacitance
of the JJ per unit area; 
~$\kappa =\hbar^2c^2/(16\pi de^2)$,
with ~$c ,\,\,  d$~
standing for the speed of light, and the magnetic thickness of the JJ (given as the sum of
the geometrical thickness ~$d_b$~ of the barrier and of the twice of the
London penetration length \cite{BARONE}),
respectively. We take into account the Josephson
relation

\begin{eqnarray}
\displaystyle
V={\hbar \over 2e}\dot{\varphi}
\label{9}  
\end{eqnarray}
\noindent
between ~$\varphi ({\bf x},t)$~ and the voltage ~$V=V({\bf x},t)$~
across the JJ, and find 

\begin{eqnarray}
\displaystyle
H=H_{\xi} +H_{EM} +H_J =
\int d^3x[{\rho \dot{\xi}^2\over 2} 
+ {\rho \omega_0^2\over 2} \xi^2 +
{\rho v_g^2\over 2} (\nabla \xi)^2] +
\nonumber
\\
\phantom{XXXXX}
\label{102}  
\\
+\int d^2x[{C\over 2}
\left({\hbar \over 2e} \right)^2\dot{\varphi}^2
+ {\kappa \over 2}({\bf \nabla}\varphi)^2
+E_J^{(0)}(\varphi)(1 + b\xi) ],
\nonumber
\end{eqnarray}
\noindent
where the integrations ~$\int d^3x$~ and ~$\int d^2x ...$~ 
are performed 
over the bulk and over the JJ barrier
area, respectively. 
The above form yields equations of motion
of the joint dynamics of ~$\varphi$~  and ~$\xi$. 
In this paper we will 
concentrate on the effect of small oscillations.

The Josephson equation following
from eq.(\ref{102}) is 

\begin{eqnarray}
\displaystyle
\ddot{\varphi} + \gamma \dot{\varphi}
-\overline{c}^2\nabla^2\varphi +\nu 
J_0(\varphi)(1+b\xi)=0, \quad
 \nu={ 2e\over \hbar C},
\label{2}  
\end{eqnarray}
\noindent
where all the variables are taken 
at the JJ plane; 
~$\overline{c}=c/\sqrt{4\pi dC}$~ stands
for the effective speed of light in the JJ; 
the term ~$\gamma \dot{\varphi}$~ describes
the effect of dissipation in the resistively
shunted JJ \cite{BARONE}, and ~$\gamma =1/(\rho_s C)$,
with ~$\rho^{-1}_s$~ being the shunt 
conductance per unit area of 
the JJ. 

The equation for ~$\xi$~ also follows from
eq.(\ref{102}) as 

\begin{eqnarray}
\displaystyle
\ddot{\xi} + \omega_0^2\xi -v^2_g\nabla^2\xi
+{b\over \rho}E^{(0)}_J(\varphi)\delta (z) =0,
\label{20}  
\end{eqnarray}
\noindent
where we have specified the JJ barrier 
as the plane ~$z=0$. Note that, while the Laplace 
operator 
~$\nabla^2={\partial^2 \over \partial x^2}
+ {\partial^2 \over \partial y^2}$~ 
in eq.(\ref{2}) acts along the JJ plane, the
Laplace operator 
~$\nabla^2={\partial^2 \over \partial x^2}
+ {\partial^2 \over \partial y^2} +
{\partial^2 \over \partial z^2}$ in eq.(\ref{20})
acts in the bulk.

Equilibrium 
currents correspond to a non-trivial static solution
of eqs.(\ref{2}), (\ref{20}) which in the static case
become

\begin{eqnarray}
\displaystyle
-\overline{c}^2\nabla^2\varphi_0 +\nu 
J_0(\varphi_0)(1+ b\xi_0(z=0) )=0,
\label{5}  
\\
\phantom{XXXX}
\nonumber
\\
\displaystyle
\omega_0^2\xi_0 -v^2_g\nabla^2\xi_0
+{b\over \rho}E^{(0)}_J(\varphi_0)\delta (z) =0,
\label{50}  
\end{eqnarray}
\noindent
where ~$\xi_0(z=0)$~ stands for the solution
of eq.(\ref{50}) taken at the JJ plane ~$z=0$.
In the long wave limit, the solution of
eq.(\ref{50}) can easily be found and substituted
into eq.(\ref{5}), so that
it acquires the form

\begin{eqnarray}
\displaystyle
-\overline{c}^2\nabla^2\varphi_0 +\nu 
J_0(\varphi_0)[1 - gE^{(0)}_J(\varphi_0)]=0,\,\,\,
g={b^2\over 2\rho |\omega_0 v_g|}.
\label{500}  
\end{eqnarray}
\noindent
It can now be seen that the interaction with 
phonons modifies the equilibrium Josephson
current, so that the effective current
becomes ~$\tilde{J}(\varphi)=J_0(\varphi)
[1-gE^{(0)}_J(\varphi)]$. The resulting
current may 
significantly deviate from ~$J_0(\varphi)$,
if the coefficient ~$g$~ is large enough.
This may occur if $b$ is anomalously
large, or the phonon mode exhibits
soft-mode behavior. 
In this paper
we will not consider such possibilities,
and therefore it should be clearly
shown that the dynamical effect
under consideration does not imply
strong deviations 
of the current-phase relation
(CPR) from the bare one ~$J_0(\varphi)$,
even though the values of $b$ are
taken large as discussed above (see
below eq.(\ref{1})).
To this end, we assume, for a while, that
the current
obeys the standard CPR 
~$J_0=J_c \sin \varphi$ \cite{BARONE}.
Then, 
the effective
current  becomes
~$J=J_c(1-g\hbar J_c/(2e))\sin \varphi + 
(g\hbar J^2_c/(4e)) \sin 2\varphi$.
We require that the second term,
which is ~$\sim \sin 2\varphi$,
is much smaller than the main one
~$\sim \sin \varphi$, that is,

\begin{eqnarray}
\displaystyle
\tilde{g}={g\hbar J_c\over 4 e}=
{b^2\hbar J_c\over 8e\rho |\omega_0 v_g|}\ll 1\, .
\label{5000}  
\end{eqnarray}
\noindent
Later we will show
that for typical values of the parameters,
~$ \tilde{g}\approx 10^{-2} - 10^{-3}$,
so that eq.(\ref{50}) as well as the
term ~$\sim \xi_0$~ in eq.(\ref{5})
can be safely ignored. 

Here we do not specify mechanisms
which may result 
in deviations of ~$J_0(\varphi)$~
from the standard CPR, and,
as a possible consequence, in
spontaneous Josephson currents
(see, e.g., \cite{TRBS,CPR}).
Our goal here is studying
implications of the presence of such
currents on the process under consideration.
In fact, the equilibrium currents
may occur due to some trapped integer vortices or
external magnetic field as well.

Below we will consider coupling between 
small oscillations of ~$\varphi$, which
constitute small deviations from the 
equilibrium, and ~$\xi$.
We introduce small deviations 
~$\psi ({\bf x}, t)$~ of ~$\varphi$~
from ~$\varphi_0$~ as 

\begin{eqnarray}
\varphi ({\bf x}, t)=
\varphi_0({\bf x}) + \psi ({\bf x}, t).
\label{4}
\end{eqnarray}
\noindent
The variable ~$\xi$~ should also be represented
as the sum of the static solution ~$\xi_0$~ 
of eq.(\ref{50})
and the oscillatory part. We will, however, ignore
the static part, provided eq.(\ref{5000}) holds.

The form (\ref{4}) can be employed
in eq.(\ref{2}) in order to obtain
a linearized equation for 
~$\psi$. Before, however, we will show
explicitly that, as long as the spontaneous
currents are present, the variables
~$\psi$~ and ~$\xi$~ are coupled linearly.  
Indeed, 
substituting (\ref{4}) into eq.(\ref{102})
and selecting the lowest order terms describing
the interaction between ~$\psi$~ and ~$\xi$,
we find the interaction energy as

\begin{eqnarray}
\displaystyle
H_{int}= \int d^2x
 {\hbar b\over 2e}J_0(\varphi_0)\psi \xi\,\, ,
\label{4_2}
\end{eqnarray}
\noindent
where ~$\xi$~ is taken at the JJ plane,
and the definition of
~$J_0(\varphi)$~ (see below eq.(\ref{1_1}))
has been employed.
Eq.(\ref{4_2}) indicates 
that the spontaneous (equilibrium) currents
~$ J_0(\varphi_0)$~ induce
the first order coupling between
small oscillations  ~$\psi$~  
and ~$\xi$.

It is worth noting that, the form
(\ref{4_2}) of the coupling
between the Josephson phase oscillations ~$\psi$~
and a phonon variable ~$\xi$~ is the lowest
possible coupling as long as ~$\xi$~ describes a
non-polar (IR-inactive) mode. 
The situation is very much different
in the case of the IR-active mode.
Indeed, the IR-active phonons induce
electric fields which interact with the Josephson
phase directly as indicated by the Josephson relation
(\ref{9}). Such a coupling will lead to the 
dissipation of the Josephson plasma oscillations
due to transferring its energy to the IR-active
phonons. This effect has been studied quite well
\cite{KINDER}, and it may serve as a reference 
for assessing a significance of the IR-absorption
in the case of the non-polar phonons.
In order to make a proper comparison, we introduce
some polar mode.
It is a vector displacement field ~${\bf
\xi}'({\bf x},t)$~ in the barrier, which corresponds 
to the electric polarization density
~$ea^{-3}\bf \xi'$~ (taken per
unit cell
of the volume ~$a^3$).
Then,
the interaction energy between the a.c. voltage ~$V$~
and  ~${\bf \xi}'({\bf x},t)$~ is simply the polarization
energy

\begin{eqnarray}
\displaystyle
H'_{int}= -{e\over a^3}\int d^3x
 E_z(x,y,t) \xi'_z(x,y,z,t),
\label{110}
\end{eqnarray}
\noindent
where ~$E_z$~ is the electric field in the barrier,
which is taken to have only one component 
along the normal to the barrier and is assumed
to be uniform along this
direction. Then, one finds
~$ E_z=V/d_b$~, and ~$E_z$~ can be related
to the phase oscillations by means
of the Josephson relation (\ref{9}). 
The integration is carried out inside
the insulating barrier of the width ~$d_b$.
Thus, assuming that
the barrier is thin, so that ~$\xi'_z$~ can be taken
uniform along the $z$-axis inside the barrier, 
we find eq.(\ref{110}) as   

\begin{eqnarray}
\displaystyle
H'_{int}= -{\hbar \over 2a^3}
 \int dxdy 
 \dot{\psi}(x,y,t)\xi'_z(x,y,t).
\label{120}
\end{eqnarray}
\noindent
Thus, the polar mode interacts
directly with the phase oscillations regardless
of the presence of the stationary currents. 
Later we will employ the form (\ref{120})
as a reference for assessing the strength of the 
IR-absorption arising in the case of the
non-polar mode described by
eq.(\ref{4_2}). 

We consider two types of processes initiated
by the term (\ref{4_2}): i) oscillations ~$\xi$~ can excite
the JJ plasma waves ~$\psi$ at the same frequency, 
in accordance with
the  spatial profile
of ~$ J_0(\varphi_0({\bf x}))$;
ii) the incoming electromagnetic radiation
excites the phase oscillations which convert
into the oscillations ~$\xi$~ directly
at the same frequency. This constitutes the
IR-absorption by the originally IR-inactive mode.
The intensity of such an absorption is determined
by ~$ J_0(\varphi_0({\bf x}))$.
Accordingly, the stationary current distribution 
~$ J_0(\varphi_0({\bf x}))$~
can be
inferred.

\section{Direct conversion of non-polar 
phonons into the Josephson plasma oscillations}

In this section, we consider the process i). 
To be specific, we choose  ~$\xi(x,y,z, t)$~
as the plane wave 
~$A_{\omega}
\exp [i(\omega t - {\bf kx})] +c.c.$,
characterized by the frequency ~$\omega$~ and 
by the wave vector ~${\bf k}=(k_x,k_y, k_z)$~ as well as 
by some constant amplitude ~$ A_{\omega}$.
The junction plane is taken at ~$z=0$, and
the insulating layer (the barrier) is taken
as being of zero thickness. This approximation
can be justified, if the actual barrier thickness ~$d_b$~
(which is about few nm) is much
smaller than ~$1/k_z$, so that ~$\xi $~ does not
vary significantly inside the barrier along the normal. 
Hence, in what follows 
we will employ the form  

\begin{eqnarray}
\xi = A_{\omega}
\exp [i(\omega t - k_xx-k_yy)] + c.c.
\label{3}
\end{eqnarray}
\noindent
at the location of the barrier 
in eqs.(\ref{2}),(\ref{4_2}). Performing, then, the
linearization of eq.(\ref{2}),
we obtain 

\begin{eqnarray}
\displaystyle
(-\Omega^2
+\omega^2_J({\bf x}))\psi_{\omega}({\bf x})
-\overline{c}^2\nabla^2\psi_{\omega}({\bf x})
 +b\nu 
J_0(\varphi_0({\bf x})) A_{\omega}
{\rm e}^{-i(k_xx+k_yy)} =0,
\label{6}  
\end{eqnarray}
\noindent
where we have introduced the notation
~$\psi_{\omega}({\bf x})$~ for the time
independent amplitude of
~$\psi ({\bf x}, t)= \psi_{\omega}({\bf x})
{\rm e}^{i\omega t} + c.c. $; 
~$\Omega^2=\omega^2 - i\gamma \omega$; and

\begin{eqnarray}
\displaystyle
\omega^2_J({\bf x})=\nu {\partial J_0(\varphi_0)
\over \partial \varphi_0 }
\label{7}  
\end{eqnarray}
\noindent
is the square of the local Josephson plasma
frequency. 

Eq.(\ref{6}) describes the
electromagnetic linear response of the Josephson phase on the
phonon field (\ref{3}) \cite{COM2}.
It is important to realize that the resulting
a.c. Josephson voltage (\ref{9}) turns out 
to be given by the equilibrium Josephson currents
~$J_0(\varphi_0)$. This feature can be employed 
for the imaging of such currents.
It should, however, be noted that 
the non-uniformity of ~$\omega^2_J({\bf x})$~ (\ref{7})
complicates solving eq.(\ref{6}). In order
to avoid
this problem, we will consider the case 

\begin{eqnarray}
\displaystyle
|\Omega^2|=|\omega^2 - i\gamma \omega|
\gg \omega^2_J\,\, .
\label{71}  
\end{eqnarray}
\noindent
Taking such a limit, on one hand, allows omitting the term
~$\sim \omega^2_J$~ from eq.(\ref{6}), which simplifies
the solution significantly. 
On the other hand,
eq.(\ref{71}) implies that 
the magnitude of the produced
voltage is suppressed. A
solution of eq.(\ref{6}) for the uniform a.c. voltage
~$V(t)=V_{\omega}\exp (i\omega t) + c.c.$, where
~$ V(t)=\hbar /(2eS)\int d^2x \dot{\psi}$~ and
the integration is performed over the whole
area ~$S=LW$~ of the junction, 
with ~$L$~
and ~$W$~ standing for the length (along $x$)
and the width (along $y$) of the junction,
respectively,
is

\begin{eqnarray}
\displaystyle
V_{\omega}=Z_{\xi}(\omega) J_{0\bf k},\,\,\,
J_{0\bf k}=
\int d^2x
J_0(\varphi_0({\bf x}))
{\rm e}^{-i(k_xx+k_yy)}, \,\,\,
Z_{\xi}(\omega)=
{i b\omega A_{\omega} \over LW C \Omega^2} \, ,
\label{8}  
\end{eqnarray}
\noindent
 where ~$J_{0\bf k}$~ stands for the 2D Fourier component
of ~$J_0(\varphi_0)$~ along the GB;
an explicit expression
for ~$\nu$~ (see eq.(\ref{2})) has been
employed. 
The solution (\ref{8}) 
 shows that the uniform a.c. voltage
induced by the non-polar phonons incident on the
JJ is proportional to the spatial Fourier component
of the static Josephson current ~$J_0(\varphi_0({\bf x}))$~
at the wave vector ~($k_x,k_y$)~
(along the junction plane) of
the incident phonon field. 

Let us estimate magnitude of the voltage
(\ref{8}) in the non-resonant regime,
that is, when the condition (\ref{71})
is satisfied.
A typical extension of the current structure
in the JJ is given by the Josephson penetration length
~$L_J=\sqrt{\hbar c^2/
(8\pi e d J_c)}\approx  1\mu$m for the critical
currents  ~$J_c\approx 10^5{\rm A/cm^2}$ and
~$d\approx 3\cdot 10^{-5}$cm.
Thus, the projection of the phonon wave vector
along the junction plane (the ~$k_{x,y}$~ components)
should be 
~$k_{||}\approx L^{-1}_J\approx  
10^4 $cm$^{-1}$ in order to produce
the most efficient imaging. 
In this case, 
~$J_{0\bf k}\approx J_cL_JW$ for ~$W\ll L_J$~
and ~$L\gg L_J$.
Then, we choose
~$bA_{\omega}=10^{-2}$~ which corresponds to the maximum
relative displacement (deformation) about ~$10^{-4}$.
We also employ the values 
~$\overline{c}/c\approx 10^{-2}$~
\cite{BARONE}, yielding 
~$\omega_J=\overline{c}/L_J
\approx 10^{12}$s$^{-1}$. In order
to satisfy the condition (\ref{71}),
we choose ~$\Omega^2\approx 10\omega^2_J$~ which
corresponds to

\begin{eqnarray}
\displaystyle
|\Omega|=\sqrt{|\omega^2 - i\gamma \omega|}
\approx 3\omega_J.
\label{72}
\end{eqnarray}
\noindent
When dissipation is negligible
(large shunt resistance, so that ~$\gamma
\ll \omega_J$~ in eq.(\ref{2})), the
condition (\ref{72}) can be satisfied by 
~$\omega \geq 10^{13}$Hz. While
being in the range of the optical 
frequencies, so high ~$\omega$~
is hardly achievable for sound. 
If, however, ~$J_c$~ is in the range
$1-10$A/cm$^2$, the plasma
frequency becomes 
~$\omega_J\approx 10^9-10^{10}$s$^{-1}$,
and the condition (\ref{72}) can easily
be fulfilled for sound.
It should also be noted that a  
less demanding requirement for the
sound can be imposed for the
over-damped JJ, where ~$\gamma \gg \omega_J$~
in eq.(\ref{2}). In this case, already ~$
\omega^2_J/\gamma \leq \omega \ll \omega_J $~ 
fulfills the condition
(\ref{72}). This allows one to
choose  ~$\omega $~ which is much less
than the above values. For example, for ~$\gamma =
10^{14}$s$^{-1}$ and ~$\omega_J=10^{12}$s$^{-1}$, 
the condition (\ref{72})
is satisfied by ~$\omega=100$GHz, which is on
the upper limit for the hypersound techniques
\cite{HYPER}. Finally, setting the length of the
junction ~$L\approx 10L_J$, we obtain
the value ~$V_{\omega}\approx 10^{-8}$V
in eq.(\ref{8}) \cite{COM3}. 

Let us also justify the validity of 
the approximation neglecting eq.(\ref{50}).
As discussed above, taking eq.(\ref{50})
into account may result in significant
deviations of the CPR from the 
standard one. This, however, does not occur
if the condition (\ref{5000}) holds.
Employing the above
values in eq.(\ref{5000}) and
choosing ~$\omega_0\approx \omega =10^{12}$s$^{-1}$
and ~$v_g=10^5$cm/s, ~$\rho \approx 6$g/cm$^3$, 
we find ~$\tilde{g}\approx
3\cdot 10^{-3} \ll 1$~ as mentioned above.

\section{ IR-absorption 
induced by the static Josephson currents}

Now let us consider the case ii), that is, the possibility
of the coherent phonon field  
generation as a result
of applying a uniform a.c. voltage 
~$V_{\omega}\exp (i\omega t) +c.c.$~ across the barrier.
This voltage corresponds to the phase oscillations

\begin{eqnarray}
\displaystyle
\psi (t)={2e V_{\omega}\over i\hbar \omega}
{\rm e}^{i\omega t} + c.c.,
\label{90}  
\end{eqnarray}
\noindent
following from eq.(\ref{9}).
These phase oscillations serve as a source (in eq.(\ref{20}))
for the 
oscillations of ~$\xi $~ 
in the barrier, which are subsequently 
emitted  
into the bulk.  In order to describe this 
effect, we employ eqs.(\ref{20}),(\ref{4}), and retain
only the part ~$\sim \psi$~ 
in the last term of eq.(\ref{20}).
Accordingly, eq.(\ref{20}) becomes

\begin{eqnarray}
\displaystyle
\ddot{\xi} + \omega_0^2\xi -v^2_g\nabla^2\xi
+ \left({ bV_{\omega} J_0(\varphi_0)\over i \rho \omega}
{\rm e}^{i\omega t} + c.c.\right)\delta (z)=0,
\label{200}  
\end{eqnarray}
\noindent
 where eq.(\ref{90}) has been employed. 
Eq.(\ref{200}) describes the linear response of 
~$\xi$~ on
the external a.c. voltage. This can
be interpreted as though the displacement
field ~$\xi$~ has become polar,
so that it exhibits the IR-activity
inside the JJ barrier. As a result,
an additional dissipation of the external
radiation occurs.
The total power ~$P(\omega)$~ 
carried away by the phonons, which are
generated in the barrier and, then, are emitted into the
bulk, can be calculated by means of solving eq.(\ref{200}) 
and substituting the result into the expression
for the phonon energy flux 
density ~$-\rho v^2_g\dot{\xi}{\bf \nabla}
\xi$. Averaging this flux 
over time and integrating over the JJ area, 
we find

\begin{eqnarray}
\displaystyle
P(\omega)=R^{-1}(\omega)|V_{\omega}|^2,\quad
R^{-1}(\omega)={b^2\over |\omega | \rho v_g^2}
\int {d^2q\over (2\pi )^2}{|\int d^2x {\rm e^{i\bf qx}} 
J_0(\varphi_0)|^2\over \sqrt{v_g^{-2}(\omega^2 
-\omega^2_0)-q^2}},
\label{201}  
\end{eqnarray}
\noindent
 where ~$\omega > \omega_0$~ (for ~$v^2_g >0$~); 
and the integration
is performed over the range ~$|q|<\sqrt{ v_g^{-2}(\omega^2 
-\omega^2_0)}$. The above result indicates that the
frequency dependence of the absorbed power is
determined by the spatial profile of the stationary 
current distribution.
Thus, measuring the IR-absorption
can be employed for obtaining
information about ~$J_0(\varphi_0)$.
The quantity ~$R^{-1}(\omega)$~ carries an obvious
meaning of the a.c. conductance due to the
phonons.

It is useful to compare the absorption
intensity (\ref{201}), caused by the emission of
the {\it non-polar phonons}, 
with the intensity in the
case when the phonons are polar, and
the coupling is represented by eq.(\ref{120}).
Such a comparison helps evaluating a
significance of the effect under
consideration.
This is especially important because
the technique of the IR-absorption 
in the JJ is well established \cite{KINDER}, and
therefore can be employed for detecting the effect
described above.
Thus, we assume that the displacement fields
in the both cases are the same ( ~$\xi=\xi'_z$).
In the case (\ref{120}), the last
term in eq.(\ref{200}) should be replaced by
 ~$eV/(\rho a^3)\delta (z)$. The corresponding emission
intensity ~$P'(\omega)$~ can be calculated similarly
to how eq.(\ref{201}) has been
obtained. Comparing
it with (\ref{201}), we find

\begin{eqnarray}
\displaystyle
{P(\omega) \over P'(\omega)}
\approx \left({bJ_ca^3\over e\omega }\right)^2,
\label{13}
\end{eqnarray}
\noindent
where we have estimated the integrals in
eq.(\ref{201}) as ~$\int d^2q|\int d^2x
J_0\exp (i{\bf qx})|^2\approx L_JW|J_c|^2$,
and has selected the emission area, from
which ~$P'(\omega)$~ is collected, as
~$ L_JW$. Obviously, in the case
represented by eq.(\ref{201}), ~$ L_JW$~
is the effective area occupied by ~$J_0(\varphi_0)$.
Taking ~$\omega =10^{12}$s$^{-1}$,
~$a=0.4$nm, ~$b=10^2/a$ and ~$J_c=10^5$A/cm$^2$,
we find the ratio (\ref{13}) as ~$\approx 10^{-2}$. For
~$\omega =10^{11}$s$^{-1}$ this ratio 
becomes close to 1. 
 Thus, the IR absorption per unit area
occupied by the equilibrium Josephson
currents can be comparable to the absorption due to the 
IR-active mode. This conclusion makes
 detecting the effect described above 
by means of employing the technique \cite{KINDER}
quite feasible.

\section{Imaging of the stationary Josephson 
currents by sound}

The effect discussed in Sec.III gives
a principal possibility of imaging the 
currents ~$J_0(\varphi_0({\bf x}))$~ 
by means of detecting
 uniform component of the a.c. voltage
produced as a result of subjecting the
barrier to a coherent time dependent
field ~$\xi$. In this case,
eq.(\ref{8}) relates such a voltage to
a $single$ spatial Fourier component of 
~$J_0(\varphi_0({\bf x}))$.
The drawback of this is
that, in order to keep  such a relation 
valid, the frequency of the incoming
phonon field should be high enough.
As we will discuss below, this limitation
can be circumvented by employing the
reverse effect discussed in Sec.IV, that is, 
a coherent emission of the non-polar phonons induced
by external electro-magnetic 
radiation. 
In this case, measuring the phonon field ~$\xi$~
(in, e.g., far
zone) which obeys eq.(\ref{200}) would allow
to achieve the same goal.     
We will discuss
this in detail for sound, so that
detecting a dynamical mechanical stress in the
far zone could be employed for the imaging
of ~$J_0(\varphi_0({\bf x}))$. 

In what follows the role of ~$\xi$~
is played by the strain tensor ~$u_{ij}$.
We employ a simplest model of the
dilatation interaction between 
~$u_{ij}$~ and the Josephson
currents, so that we take ~$\xi
=u_{ii}$, and
describe the dynamics in the bulk
within the isotropic
medium approximation. 
Thus, the energy explicitly
dependent on ~$u_{ij}$~ is

\begin{eqnarray}
\displaystyle
H_d=\int d^3x[{\rho \dot{u}_i^2\over 2} 
+ {\lambda \over 2} u_{ii}^2 + \mu u_{ij}^2
+\eta \psi\,\, u_{ii}] ,\quad
\eta ={\hbar b J_0(\varphi_0)\over 2e}
\delta (z)\, ,
\label{14}  
\end{eqnarray}
\noindent
where 
 ~$\lambda ,\,\, \mu$~ are the Lame
coefficients; ~$\rho$~ stands for the
total density; the source term ~$\sim \eta $~
is due to ~$H_{int}$~ (\ref{4_2}).

Let us assume that 
small uniform phase oscillations  ~$\psi (t)$~
(\ref{90})
are imposed externally by the incoming 
microwave radiation \cite{COM4}. We represent
the deformation field as ~$u_i=\nabla_i u$, 
which determines ~$u_{ij}=(\nabla_iu_j +
\nabla_ju_i)/2$, with
~$u$~ being some scalar field. We note that
 the transverse part of the deformations is decoupled
in this simplified approach. 
Then,
the equation for ~$u$~, following from
eq. (\ref{14}), is

\begin{eqnarray}
\displaystyle
\ddot{u} -v_s^2\nabla^2u 
={2e \eta \over i\hbar \omega \rho}
V_{\omega}{\rm e}^{i\omega t} + c.c.\,\, ,
\label{15}  
\end{eqnarray}
\noindent
where ~$v_s=\sqrt{(\lambda + 2\mu)/\rho}$~ 
stands for the speed of the longitudinal sound;
eq.(\ref{90}) has been employed;
and no dissipation of sound is considered.
Solution of eq.(\ref{15})
 can easily be 
found in terms of the Fourier harmonics
~$u=u_{\bf q}
{\rm e}^{i\omega t -i{\bf qx}} + c.c.$~ as

\begin{eqnarray}
\displaystyle
u_{\bf q}= {2e V_{\omega}\over i\hbar \omega \rho
(v_s^2q^2 - \omega^2)} \eta_{\bf q}\, ,
\label{150}  
\end{eqnarray}
\noindent
where ~$\eta_{\bf q}=\int d^3x \exp (i{\bf qx})\eta 
({\bf x})$~ denotes the 3D Fourier harmonic
of ~$\eta$~ in eq.(\ref{14}), and ~$V_{\omega}$~ 
is taken uniform along the junction.  
As represented,
eq.(\ref{150}) is valid for 
an arbitrary configuration
of the source ~$\eta$. Thus, it is not limited by
the specific form (\ref{14}) corresponding
to the plane junction.

Let us find the $\omega$-harmonic of the pressure  field 
~$\sigma = \rho v^2_s u_{ii}=\rho v^2_s \nabla^2u$~ 
far from the junction in the effective 2D geometry. That is,
we assume uniformity of all the quantities in the
$y$ direction corresponding to the thickness
$W$ of a slab containing the JJ plane ~$z=0$. The other
two directions ~$x,\,\, z$~ are unlimited. 
The location of the spontaneous currents 
determined by the solution
of eq.(\ref{5}) is
restricted to some region of a typical 
size $L_J$ (along ~$x$~) and ~$W$~ (along $y$). 
The distance $r$ (along
~$x,\,\, z$~ directions) at which the
stress is detected is considered to be
much larger than ~$L_J$~ and ~$W$. Performing
the inverse Fourier transform of eq.(\ref{150})
in the far zone, we find the 
~$\omega$-harmonic of the pressure ~$\sigma (x,z,t)
=\sigma_{\omega}(r, \theta)\exp (i\omega t) + c.c.$~  as

\begin{eqnarray}
\displaystyle
\sigma_{\omega} (r,\theta)=
{e q^{3/2}_{\omega} V_{\omega}\over 
\sqrt{2\pi }\hbar \omega} {\sqrt{i}{\rm e}^{-iqr}\over
\sqrt{r}} \eta_{\bf q} ,
\label{16} 
\end{eqnarray}
\noindent
where the notation ~$q_{\omega}=\omega/v_s$~
has been introduced, and ~${\bf q}=
q_{\omega}{\bf n}$, with ~$\bf n$~ 
standing for the unit vector in the 
direction of observation from the
location of the currents, that is, ~${\bf r}= (x,z)$
and ~${\bf r}=r{\bf n}$, with ~$n_x=\sin \theta ,\,\,\,
n_z=\cos \theta $~ given by the angle of observation
~$\theta$ with respect to the axis ~$z$. 
Eq.(\ref{16}) can be employed 
for imaging of the spontaneous currents
by means of detecting the mechanical
pressure (stress) at the frequency
of the external a.c. voltage 
far from the junction.

For the purpose of estimating the
magnitude of the stress, we assume that 
the equilibrium Josephson current is uniform in the
~$y$-direction. Then, after employing 
the explicit expression
for ~$\eta$~ (\ref{14}) in eq.(\ref{16}),
we find 

\begin{eqnarray}
\displaystyle
\sigma_{\omega} (r, \theta )=
{b q^{3/2}_{\omega} V_{\omega}\over
2\sqrt{2\pi }\omega} {\sqrt{i}{\rm e}^{-iqr}\over
\sqrt{r}}  
\int dx{\rm e}^{i(q_{\omega}\sin \theta) x}
J_0(\varphi_0 (x)).
\label{161}
\end{eqnarray}
\noindent
This expression shows that, if the
stress amplitude in the far zone
is known for a sufficient range of the angles
of observation ~$\theta$~ and frequencies
~$\omega$, the 
stationary (spontaneous) Josephson
current distribution
can be restored by means of the
spatial inverse Fourier transform.
Later we will estimate
the expected values of ~$\sigma$.

Now let us obtain an expression for the
total electric power dissipated due to 
the emission of sound (see eq.(\ref{201})
for optical phonons). 
As we will see,
this power demonstrates 
a universal behavior ~$\sim \omega^3$~
as ~$\omega \to 0$~ as long as 
the current distribution  ~$J_0$~
is characterized by some finite vorticity.

The dissipated power ~$P(\omega)$ equals to
the total flux of energy

\begin{eqnarray}
\displaystyle
P(\omega)=- rW\int d{\bf n} \sigma  
{\bf \nabla}\dot{u}
\label{17}
\end{eqnarray}
\noindent
carried away by the
sound. Here the surface 
integral of the energy flux density is
taken over the cylinder surface of the 
radius ~$r\gg L_J$ and of the height ~$W$.
A substitution of eqs.(\ref{150}), (\ref{161})
yields

\begin{eqnarray}
\displaystyle
P(\omega)=
R^{-1}(\omega)|V_{\omega}|^2,\quad
R^{-1}(\omega)= 
{Wb^2\omega
\over 4\pi \rho v_s^4} \int d \theta \,\,
\left|\int dx\,\,{\rm e}^{i(q_{\omega}\sin \theta) x}
J_0(\varphi_0 (x))\right|^2. 
\label{18}
\end{eqnarray}
\noindent
It is important to note that the frequency dependence
of ~$R^{-1}(\omega)$~ turns out to be sensitive
to the net vorticity. Indeed, let us consider the case
when the JJ contains some vortex characterized
by the phase variation ~$\varphi_0(x)$~ along the junction,
so that the phase difference on ~$x=\pm \infty$~ is
~$\delta \varphi_0 \neq 0$.
Then, taking into account 
 eq.(\ref{5}) in eq.(\ref{18}), we obtain 

\begin{eqnarray}
\displaystyle
R^{-1}(\omega)=R^{-1}_0\omega^3, \quad
R^{-1}_0={\hbar^2c^4 b^2W
|\delta \varphi_0|^2
\over 256 \pi^2 e^2 v_s^6d^2\rho}\,\, ,
\label{21}
\end{eqnarray}
\noindent
where
the wavelength ~$2\pi /q$~ of the
emitted sound is taken longer than the Josephson
length ~$L_J$ ( along which the variation
of the phase typically takes place), 
so that the Fourier transform
of ~$J_0$~ was taken as 

\begin{eqnarray}
\displaystyle
J_{0q}={ \overline{c}^2 \over  \nu}
iq_{\omega}\sin\theta \int^{\infty}_{-\infty}
 dx {\rm e}^{-i(q_{\omega}\sin\theta)x}
{\partial \varphi_0(x)
\over \partial x}\approx { \overline{c}^2 \over  \nu}
iq_{\omega}\sin\theta \delta \varphi_0
\label{202}
\end{eqnarray}
\noindent
in eq.(\ref{18}), and we have employed the
explicit expression
for ~$\nu$~ in eq.(\ref{2}).
Thus, the emission of sound induces the specific
contribution
~$R^{-1}(\omega)\sim \omega^3(\delta \varphi_0)^2$~ 
to the a.c. conductance of the JJ
in the limit ~$\omega \to 0$.
We note that in the absence of the
vorticity (~$\delta \varphi =0$), 
eq.(\ref{202}) yields ~$J_{0q} \sim
 q^2$, and eq.(\ref{21}) will change to become
~$R^{-1}(\omega)\sim \omega^5$. This result indicates
that a presence of the spontaneous JJ currents characterized
by finite vorticity could be revealed 
in the frequency dependence (\ref{21}), unless, of course,
the low frequency quasiparticle effects dominate
the JJ a.c. conductance.   

Let us estimate a possible magnitude 
of the effect (\ref{21}).
For this purpose we choose the following parameters:
~$\delta \varphi_0=2\pi$; ~$W=10^{-3}$cm; ~$v_s\approx
4\cdot 10^5{\rm cm/s}$;
~$\rho = 6 {\rm g/cm^3}$;~$ d=3\cdot 10^{-5}$cm;
~$b=10^2$. This yields ~$R^{-1}=3\cdot 10^{-5}
\Omega^{-1}(\omega/{\rm GHz})^3$. 
It is important to note that the dependence
(\ref{21}) saturates at some value ~$R^{-1}_{max}$~
at frequencies larger than a frequency
~$\omega_{max}$~
corresponding to the wavelength of sound 
comparable to the Josephson length ~$L_J$.
For ~$L_J=1\mu$m, we find ~$\omega_{max}
\approx v_s/L_J=4\cdot 10^9$Hz. Accordingly,
~$R^{-1}_{max}\approx 2\cdot 10^{-3}\Omega^{-1}$. 
This corresponds to ~$ R^{-1}_{max}/(WL_J)
\approx 2\cdot 10^4 \Omega^{-1}{\rm cm}^{-2}$~
of the inverse effective shunt resistance per
unit area of the JJ.

It is worth noting that measurements of ~$R^{-1}(\omega)$~
do not allow restoring completely a spatial profile
of the currents. As eq.(\ref{18}) indicates, 
~$R^{-1}(\omega)$~ is the quantity which
is the average over the directions. Furthermore,
the quasiparticle effects may dominate the
a.c. conductance. In order
to be able to restore ~$J_0(\varphi_0({\bf x}))$, the 
mechanical stress amplitude (\ref{161})
should be determined as a function of ~$\theta$~
and ~$\omega$.
Let us estimate its magnitude at distances
~$r=10^3L_J$~ far from the location
of the currents and for ~$q\approx L_J^{-1}$. 
The voltage amplitude in
eq.(\ref{161}) should be chosen in such a way
that the linearized approach is valid. Thus,
~$eV_{\omega}\leq \hbar\omega$.  Then, the Fourier 
component of the current 
~$J_{0 q}$~ in eq.(\ref{161}) can be evaluated
as ~$ J_cL_J$, 
 where a typical value of
the critical current can be taken as 
~$J_c\approx 10^5{\rm A/cm^2}$.
Then, eq.(\ref{161}) yields
~$|\sigma_{\omega}|\approx 0.1(\omega /{\rm GHz})^{3/2}$Pa 
for ~$L_J=1\mu$m.

\section{Enhancement of the interconversion
by external magnetic field}

It is worth noting that the discussed effects
can exhibit a resonant-like behavior with respect
to applied external
magnetic field ~$H_y$~ (along
the JJ width ~$W$~). Indeed, an external
field greater than the lower
critical field ~$H_{c1}\approx \Phi_0/(L_J d)$~
\cite{BARONE}, where
~$\Phi_0=hc/2e$~ is the magnetic flux
quantum, leads to
the periodic (along the junction
length) structure of the
Josephson currents due to the
Josephson vortices. 
This periodic structure can contribute
constructively to the induced voltage
(\ref{8}) and to the stress field (\ref{161}),
if the matching conditions are fulfilled.
As a result, a significant enhancement of
the corresponding values should be 
anticipated.
Let discuss this.

In the presence of ~$H_y>H_{c1}$~,
the Josephson current will change
its sign on the length ~$L_H\approx
\Phi_0/(H_yd) < L_J $~ \cite{BARONE}.
Thus, ~$L_H$~ describes the spatial period of
the main harmonic of the current.
Accordingly, as
follows from eqs.(\ref{8}),
the induced voltage will exhibit
the resonant increase when the wave
vector ~$k_{||}$~ (along the JJ barrier )
of the incoming deformation field
becomes equal to ~$2\pi/L_{H}$, that is,
for 

\begin{eqnarray}
\displaystyle
H_y\approx {\Phi_0 k_{||}\over 2\pi  d}\,\, .
\label{H2}
\end{eqnarray}
\noindent
In this case, provided the vortex lattice is
ideal, the Fourier component of the current
becomes ~$J_{0\bf k}\approx J_cWL$~ in eq.(\ref{8}).
This implies the factor of ~$L/L_J \gg 1$~
increase of the voltage magnitude if compared
with the case of a single vortex considered in Sec.III.

A similar situation occurs for the 
mechanical stress (\ref{161}) generated by the
incoming a.c. voltage. 
This stress
will exhibit a resonant
increase at certain angle of observation
~$\theta$~ given by the matching condition
~$2\pi/L_H=( \omega \sin \theta )/v_s $.
Thus, for given frequency ~$\omega$~ 
of the external radiation and
the angle ~$\theta$,
the stress amplitude
will resonantly increase at the value

\begin{eqnarray}
\displaystyle
H_y\approx {\Phi_0 \omega \sin \theta\over2\pi  v_s d}
\,\, .
\label{H1}
\end{eqnarray}
\noindent
For 
long JJ \cite{BARONE}, we find from
eq.(\ref{161})
that ~$\sigma_{\omega}$~ will increase
in proportion to the total length ~$L$~ of the
JJ. This also constitutes the factors of ~$L/L_J \gg 1$~
increase of the 
stress amplitude (\ref{161}) in the given direction,
if compared with
the above estimates for the case of a single
Josephson vortex.

\section{Discussion and conclusion}
The interconversion effect discussed 
above can, in principle,
be expected to occur in traditional superconductors
as well. Deformations should modulate the 
width ~$d_b$~ of the JJ barrier. Accordingly,
the tunneling matrix element will change.
In the superconducting state, 
this leads to the term (\ref{4_2}), where 
the coefficient ~$b$~ is determined by the
kinematics of the one electron tunneling. 
Assuming the exponential form
~$\sim \exp (-d_b/\lambda)$~ for the matrix element, 
where ~$\lambda$~ stands for a typical overlap
length of the electronic states on the both
sides of the JJ, 
the estimate ~$b\approx d_b/\lambda$~
follows, if the dominant effect of the
strain is the change ~$d_b \to d_b(1+ u_{zz})$.
Obviously, in order to have ~$b$~ as
large as chosen above ~$b\approx 10^2$, 
the critical current
~$J_c\sim \exp (-d_b/\lambda)$~ should
be practically zero. Thus, in the traditional
JJ, ~$b\approx 1-10$, which lowers the above
estimates of (\ref{8}) and (\ref{161}) significantly.
On the contrary,
in the JJ made of the high-$T_c$ materials,
the nature of the term (\ref{4_2}) is expected 
to be completely different.
The proximity of the barrier to
the superconducting state should 
result in anomalously
strong sensitivity of the barrier to 
any external factor which shifts its equilibrium.
In our estimates of ~$b$~ for the high-$T_c$ JJ,
we have relied on the indirect evidence - the
extremely strong sensitivity of the Josephson
critical current to the misorientation angle
in the GB JJ. As suggested in Ref. \cite{GB2},
the mechanical strain 
is the most plausible cause of this.

As discussed in Introduction, the mechanical strain
(stress)
may have no significant {\it direct} effect on the barrier
transport
properties, if its role is reduced to
only controlling the depletion
of oxygen at the GB during the preparation of the JJ
at high temperatures. If so, in the 
superconducting state, the external
stress will produce no substantial modulations
of the critical current, and the value
of ~$b$~ will not be much
different from the estimate for the traditional
JJ. Thus, no the strong interconversion
discussed in this paper is to be observed under
such circumstances. 
Accordingly, either observing or not
observing the discussed effect will help
elucidate the role of the mechanical
strain in suppressing the Josephson
current. If the effect is observed, this
will be a very strong indication favoring
the concept of proximity of the barrier to
the superconducting state \cite{GB3,QCL}.     

If being strong enough, 
the dynamical effect proposed above,
opens up a possibility for the 
high resolution imaging of the
equilibrium Josephson currents
in the high-$T_c$ materials. This is
especially important for revealing
the nature of the superconducting
OP at the GB, and the mechanisms of
the TRSB. The resolution is limited
by the wavelength of the sound.
Thus, for the frequencies higher
than 10GHz, the resolution becomes
better than 1$\mu$m. 

In summary, we have suggested that in the Josephson
junction, in which the critical current is sensitive
to the mechanical (optical) 
deformations, a direct conversion
of the non-polar phonons into the electromagnetic 
Josephson oscillations as well as the reverse process
can occur, if this JJ contains spontaneous currents.
We conjecture 
that these effects should be especially strong
in the high $T_c$-materials due to the proximity
of the insulating layer of the JJ to
the superconducting state. 
We propose that the effect of the interconversion can
be employed as a crucial test for the role
of deformations in the superconducting properties
of the GB JJ made of the high-$T_c$ materials.  
This effect can be used for imaging of the
Josephson current spatial distributions.  

\acknowledgements
This work was partially
supported by the CUNY Research Program Grant.


\begin{references}
\bibitem{COM2}
Strictly speaking, the form of the phonon field 
should be determined self-consistently 
from solving eqs.(\ref{2}),(\ref{20}) together
with the conditions describing
how the phonons are generated.
\bibitem{COM3}
In the resonant regime, 
when ~$\omega \approx \omega_J$~ and
the ~$Q$-factor is large, the
produced a.c. voltage can be quite large.
However, 
the non-uniformity of ~$\omega_J$ in
eq.(\ref{7}) becomes a significant factor, so
that no simple solution like eq.(\ref{8})
can be found.
\bibitem{COM4}
Strictly speaking, the form of the
phase oscillations should be determined
self-consistently from solving
eqs.(\ref{2}),(\ref{20}).
\end{references}
\end{document}